# Supporting Requirements Engineering Research that Industry Needs:

## *The Naming the Pain in Requirements Engineering Initiative*

Daniel Méndez Fernández
Technical University of Munich, Germany
daniel.mendez@tum.de

**In light** of the 40[th] jubilee of Requirements Engineering (RE), roughly 40 experts met in Switzerland to discuss where our discipline stands today [1]. 40 years after RE was acknowledged for the first time as an independent discipline in an issue of the *Transactions of Software Engineering*, a strong and (more or less) self-confident research community has emerged and contributed a variety of tools, methods, and approaches to the field. As of today, the common view is, indisputably, that "RE as a discipline is stable and respected", as pointed out by Sarah Gregory when covering the seminar in her column [1] to which articles like this one are invited to present ongoing research. However, it is also evident that after 40 years of promising research, conducting "research that industry needs" is still an ongoing challenge. Research that industry needs means research that solves industrial problems practitioners face; but do we really understand those practitioners' problems? Here, I want to recapitulate on this research challenge and outline an initiative, the Naming the Pain in Requirements Engineering Initiative, that aims at tackling this problem. Before doing so, let's briefly recapitulate the current state of Requirements Engineering Research and Practice.

## I. REQUIREMENTS ENGINEERING IS CHALLENGING – IN PRACTICE AND RESEARCH

First and foremost: Nobody can refute the importance of RE and its challenges. As an interconnected discipline, much depends on its success as many decisions – and also problems – in software development projects are rooted therein. Given its critical and challenging nature, it should thus not be surprising that project failures are still too often caused by insufficient RE. But the question is: Why are we still struggling with RE after 40 years of research? Do we lack proper scientific contributions that aim at solving practitioners' problems or do we lack proper knowledge and technology transfer between research and practice? Probably both.

For practitioners, it is often (made) difficult to share insights into their RE given its criticality for the company's competitive environment. From a researcher's perspective, it is in general very difficult to reveal clear pictures of RE in practice. The discipline is largely characterised by uncertainty and by human factors like experiences, expertise, fears, beliefs, and even politics [9]. These factors affect the choice of methods, approaches, and tools in individual practical contexts. What might work very well in one project setting might be completely alien to the needs and the culture of the next [5]. This renders it so difficult to empirically investigate the discipline. In the end, RE is a highly individual means to an end – as individual as the people working on problems and as individual as the problems they work on.

## II. CURRENT STATE OF REQUIREMENTS ENGINEERING RESEARCH

It is not surprising that the biggest challenge in RE research is to provide proper empirical figures that would demonstrate particular success factors in RE. Such factors are, however, critical determinants of what works in practice and what doesn't [2]. One consequence is that the state of empirical evidence in RE is particularly weak and that a large portion of everyday industrial practices is dominated by conventional wisdom rather than being governed by empirical evidence. Another symptom of the current disconnect between research and practice [3] is that we researchers too often keep proposing solutions to problems we have barely understood and which too often either remain unknown in practice or – worse – are blindly adopted without proper understanding of their potential benefits and shortcomings. As a consequence of the rather low practical impact of many of our contributions, the disconnect between researchers and practitioners is reinforced. The dominant role of goal-oriented requirements engineering approaches in research and the lack of such approaches in practice could be argued as one example for the current disconnect [6].

While practical relevance of research should certainly not be the sole measure of success, we still need to break this vicious cycle and foster more problem-driven research in RE. To this end, we, that is the RE community of researchers, need to close the knowledge gap on the plethora of situations practitioners face in their projects to eventually pave the road for the industrial adoption of our contributions.

### III. WE NEED PROPER THEORIES IN REQUIREMENTS ENGINEERING

Motivated by the overall situation and the current need for a stronger body of knowledge in RE, Stefan Wagner (University of Stuttgart) and I proposed the **Na**ming the **P**ain **i**n **R**equirements **E**ngineering (NaPiRE) initiative [4] back in 2012 in a workshop at the annual meeting of the International Software Engineering Research Network (ISERN[1]). The overall objective was to establish a holistic theory of RE practices and problems in industry to help guide research along specific industrial contexts and needs therein. To this end, we proposed to run NaPiRE as a bi-yearly replicated family of internationally distributed surveys where each run contributes to extend an initial theory and make it more robust over the years. We further proposed to disclose all data and resulting publications to the public to make our results transparent and reproducible, but also to support other researchers in grounding their own work on empirical data. At first, our idea received a lukewarm reception, but at the time of writing this article we number 57 researchers world-wide starting the third replication in 25 countries. The success of NaPiRE has strengthened our confidence in a shared vision and the initiative has become an effort by the community for the community.

I will now guide through an overview of selected results from the second survey replication conducted in 2014/15 with responses from 228 companies in 10 countries. I concentrate on a specific set of problems our respondents are experiencing while further details can be taken from [7]. The goal is to give the reader an overview over the diversity of problems and the importance of better understanding these problems.

### IV. NAPIRE: PROBLEMS, THEY ARE EVERYWHERE!

Figure 1 summarises the top 21 RE problems of our respondents as well as their frequency. The colour coding further shows the criticality of the problems by visualising the extent to which those problems have been experienced to lead to project failure. We can see, for example, that *incomplete requirements* constitute the most frequently stated problem directly followed by *communication flaws between the project team and the customer*, *underspecified requirements*, and *moving targets*. At the same time, we can see that *moving targets*, although ranked as the fourth most frequent problem, becomes the top priority problem when considering the project failure ratios alone. Further, when considering all problems, one may argue that many of those problems should be in scope of clear RE process models or artefact templates, that is blueprints serving as orientation on how to specify requirements or which terminology to rely on. This is interesting because many of our respondents do, in fact, use such templates. To better understand the reasons for the problems and propose fruitful solutions, we need a deeper understanding of the problems.

---

[1] http://isern.iese.de/Portal/

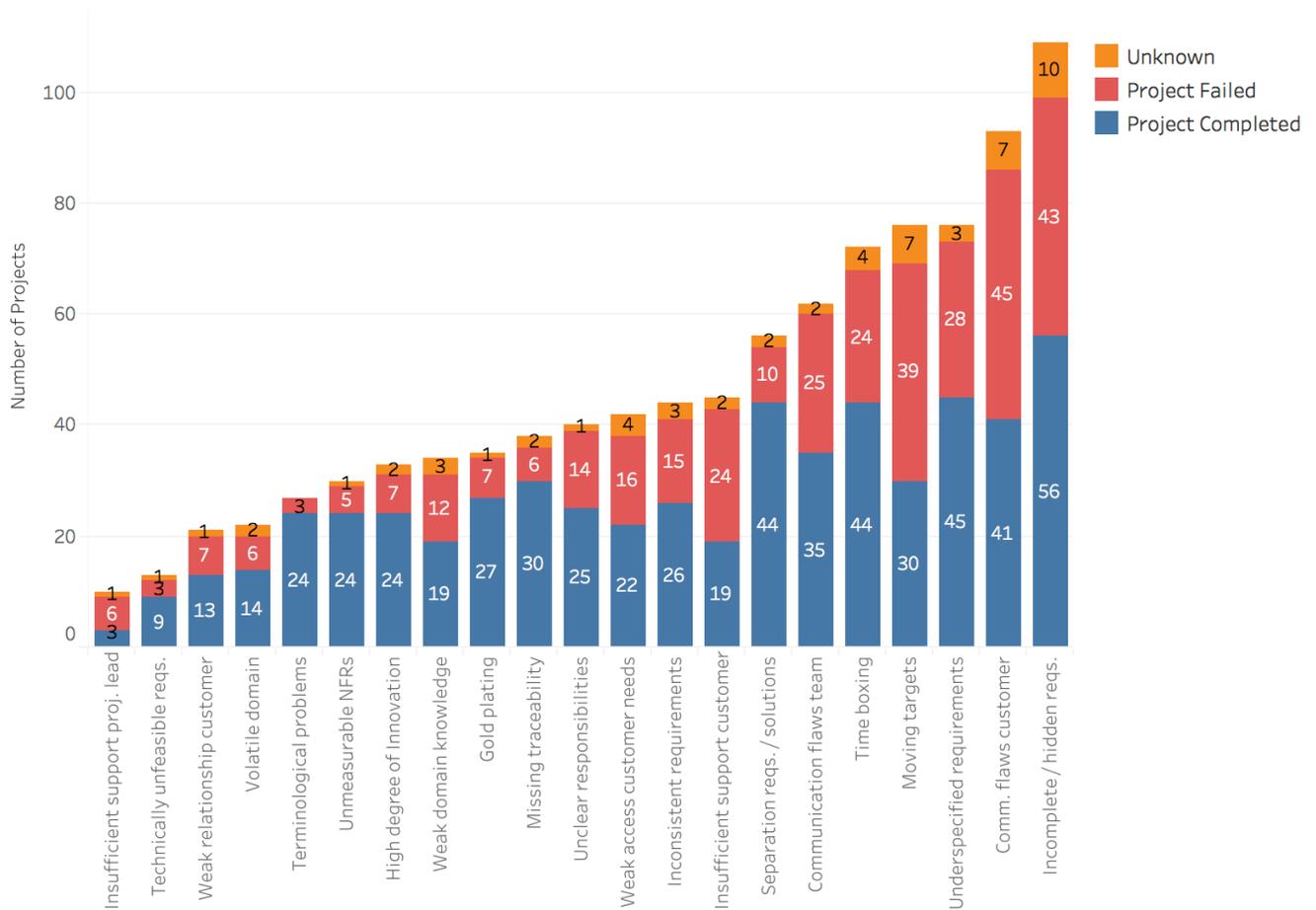

Figure 1 Key RE problems and their criticality as reported in [7].

V.  A CLOSER LOOK INTO THE PROBLEMS

The NaPiRE data captures not only practices and problems, but also root causes and consequences of the problems going beyond the simple case of project failure or not. Figure 2 visualises such an exemplary analysis for the problem *communication flaws between the project team and the customer* as a probabilistic cause-effect diagram [8] where percentages represent the share of given answers. The nearer phenomena are to the central line, the higher their probability of occurrence in projects.

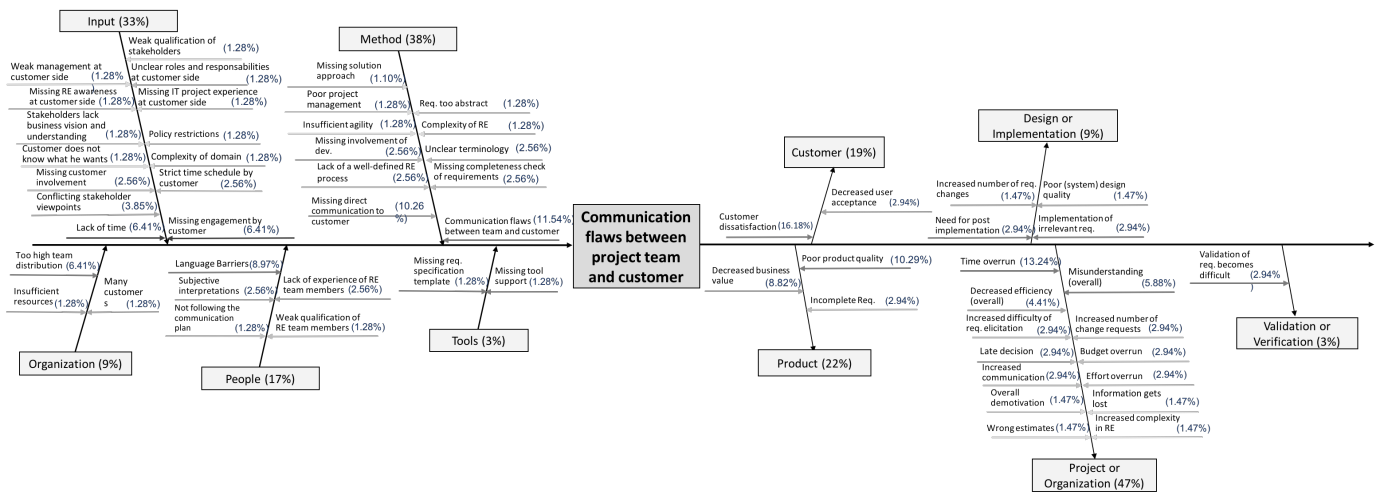

Figure 2 Causes and effects for communication flaws between the project team and the customer.

This diagram, also known as an Ishikawa diagram, shows on its right side the effects of the problem including, for instance, *poor product quality* caused by missing or incorrect features; on the left side, it shows the root-causes of the problem. It thus gives a more detailed picture going beyond problems and their apparent role in project failures. It also shows that the top 5 root causes do not indicate at all to a lack of guidance via templates, but to rather organisational and social factors (e.g., *language barriers*, *missing direct communication to customer* or *missing engagement by customer*). One natural follow-up question is: do projects applying agile practices avoid these issues given that agile practices are meant to support the communication between all involved parties?

When blocking the data according to the software process models used, we can select all responses by practitioners who apply agile practices. The outcome is summarised in Table 1 were we list the top 5 problems and group them by company size.

Table 1 Top 5 problems of companies applying agile practices as reported in [7]. Highlighted problems represent problems actually in scope of agile practices.

|  | Small Companies (1-50) | Medium-sized companies (51-250) | Large companies (>251) |
|---|---|---|---|
| # 1 Problem | Incomplete / hidden requirements | **Communication flaws between project team and customer** | Incomplete / hidden requirements |
| # 2 Problem | **Communication flaws between project team and customer** | Incomplete / hidden requirements | **Moving targets** |
| # 3 Problem | Underspecified requirements | **Communication flaws within project team** | **Communication flaws between project team and customer** |
| # 4 Problem | **Communication flaws within project team** | Separation of requirements from solutions | Time boxing |
| # 5 Problem | Time boxing | **Weak access customer needs** | Underspecified requirements |

Interestingly, all those companies still experience communication flaws between the project team and the customer. Even more, they encounter further problems that agile practices intend to solve.

## VI. What does this mean for Requirements Engineering Research?

We can see that many challenging project situations agile practices claim to address and even proclaim as their natural environment, such as moving targets, still do manifest themselves as severe problems. The article's space limitations hinder us from further reasoning about the particularities of the respondents' contexts, but the key take away of this small example does already become evident: The claims we often associate with specific solutions, like agile practices, and the expectations we have when applying them as something universal, are not clearly in tune with today's industrial reality. We all would benefit from a better understanding of industrial reality to align our research with the problems in practice.

At this moment, we are launching the third replication of NaPiRE to further explore the current state of practice in RE with a focus on the knowledge gaps such as those illustrated in the present article. This will hopefully improve the community understanding of what is *really* going on out there, and it will help us steering our research in a problem-driven manner. Until then, I cordially invite researchers to join this or similar initiatives as well as practitioners in supporting us with their answers to provide their share in producing *research that industry needs*. Details on the initiative and how to contribute can be taken from [4].

## VII. Biography

Daniel Méndez is senior research fellow at the Technical University of Munich and director of the interdisciplinary research groups at the Centre Digitisation. Bavaria. During his PhD and his subsequent habilitation, he worked on requirements engineering improvement and his research interest mainly comprises requirements engineering and empirical software engineering. Further information is available at http://www.mendezfe.org

## References

[1] S. Gregory. RE@40: Midlife Crisis or Graceful Maturity? In: *IEEE Software*, 2017.

[2] B.H.C. Cheng, and J.M. Atlee. Research Directions in Requirements Engineering. In: *Proc. Future of Software Engineering*, 2007.

[3] L. Briand, D. Bianculli, S. Nejati, F. Pastore, M. Sabetzadeh. The Case for Context-Driven Software Engineering Research : Generalizability Is Overrated. In: IEEE Software, 2017.

[4] The Naming the Pain in Requirements Engineering Initiative – www.re-survey.org

[5] D. Méndez Fernández, R. Wieringa. Improving Requirements Engineering by Artefact Orientation. In: Proc. 14th International Conference on Product-Focused Software Process Improvement, 2013.

[6] A. Mavin, P. Wilkinson, S. Teufl, H. Femmer, J. Eckhardt, J. Mund. Does Goal-Oriented Requirements Engineering Achieve its Goal? In: Proc. International Requirements Engineering Conference, 2017.

[7] D. Mendez Fernandez, S. Wagner, M. Kalinowski, M. Felderer, P. Mafra, A. Vetrò, T. Conte, M.-T. Christiansson, D. Greer, C. Lassenius, T. Männistö, M. Nayebi, M. Oivo, B. Penzenstadler, D. Pfahl, R. Prikladnicki, G. Ruhe, A. Schekelmann, S. Sen, R. Spinola, J.L. de la Vara, A. Tuzcu, R. Wieringa. Naming the Pain in Requirements Engineering: Contemporary Problems, Causes, and Effects in Practice. In: Empirical Software Engineering Journal, 2016.

[8] M. Kalinowski, E. Mendes, and G.H. Travassos. Automating and Evaluating the Use of Probabilistic Cause-Effect Diagrams to Improve Defect Causal Analysis. In: Proc. International Conference on Product Focused Software Development and Process Improvement, 2011.

[9] A. Milne, N. Maiden. Power and Politics in Requirements Engineering: A Proposed Research Agenda. In: Proc. International Requirements Engineering Conference, 2011.